\documentclass[lettersize,conference]{IEEEtran}
\usepackage[english]{babel}
\usepackage{amsmath,amsfonts,amsthm,amssymb, bbding}
\usepackage{algorithm,algorithmic}
\usepackage[caption=false,font=footnotesize]{subfig}
\usepackage{multirow}
\usepackage{stfloats}
\usepackage{url}
\usepackage{svg}
\usepackage{verbatim}
\usepackage{graphicx}
\usepackage{cite}
\usepackage{xcolor}
\usepackage{hyperref}
\hypersetup{
    colorlinks=true,
    linkcolor=blue,
    filecolor=magenta,      
    urlcolor=cyan,
    pdftitle={Overleaf Example},
    pdfpagemode=FullScreen,
    }

\begin{document}

\title{\LARGE A Parallel Monte-Carlo Tree Search-Based Metaheuristic For Optimal Fleet Composition Considering Vehicle Routing Using Branch \& Bound}

\author{T.M.J.T. Baltussen$^1$, 
        M. Goutham$^1$,
        M. Menon$^2$,
        S.G. Garrow$^2$,
        M. Santillo$^2$,
        S. Stockar$^1$}

\maketitle
\footnotetext[1]{Tren Baltussen, Mithun Goutham and Stephanie Stockar  are with the Center for Automotive Research, The Ohio State University, Columbus, OH 43212, USA. {\fontfamily{pcr}\selectfont
\{baltussen.1, goutham.1, stockar.1\}@osu.edu}}
\footnotetext[2]{Meghna Menon, Sarah Garrow and Mario Santillo are with the Ford Motor Company, Dearborn, MI 48109 USA, {\fontfamily{pcr}\selectfont
\{mmenon8, sgarrow1, msantil3\}@ford.com}}
\let\thefootnote\relax\footnotetext{This work was presented at the 2023 IEEE Intelligent Vehicles Symposium (IV) and is published under DOI: 10.1109/IV55152.2023.10186562. Copyright may be transferred without notice, after which this version may no longer be accessible.}
\begin{abstract}
Autonomous mobile robots enable increased flexibility of manufacturing systems. The design and operating strategy of such a fleet of robots requires careful consideration of both fixed and operational costs.
In this paper, a Monte-Carlo Tree Search (MCTS)-based metaheuristic is developed that guides a Branch \& Bound (B\&B) algorithm to find the globally optimal solution to the 
Fleet Size and Mix Vehicle Routing Problem with Time Windows (FSMVRPTW).
The metaheuristic and exact algorithms are implemented in a parallel hybrid optimization algorithm where the metaheuristic rapidly finds feasible solutions that provide candidate upper bounds for the B\&B algorithm.
The MCTS additionally provides a candidate fleet composition to initiate the B\&B search.
Experiments show that the proposed approach results in significant improvements in computation time and convergence to the optimal solution.

\end{abstract}

\section{Introduction}
\label{Introduction}
In the industrial sector, reconfigurable manufacturing systems are increasingly being adopted because of their ability to scale and diversify production by supporting the adaptability of process controls, functions, and operations \cite{Morgan2021IndustryMachines}.
A key enabler is the added production flexibility provided by the adoption of fleets of autonomous mobile robots (AMRs) that move material within a plant \cite{Ghelichi2021AnalyticalCenters}. 
In particular, multi-load AMRs enhance efficiency by picking up and dropping off multiple items in a single mission \cite{Yan2020AVehicles}.
The design of such a fleet is a strategic problem and involves considerable capital investment \cite{Hoff2010IndustrialRouting}. Therefore, all costs related to the acquisition and operation should be considered.
Although \cite{9993118} and \cite{8916904} have recently shown the relevance of combining vehicle routing and component design of the vehicles in the fleet, the combined vehicle routing and fleet composition has generally received insufficient attention \cite{Hoff2010IndustrialRouting}.
In this paper, the Vehicle Routing Problem with Time Windows (VRPTW) and capacity constraints on the cargo mass, volume and vehicle range is used to obtain operational costs.
The combined VRPTW with the heterogeneous fleet composition problem, is called the Fleet Size and Mix Vehicle Routing Problem with Time Windows (FSMVRPTW). This problem accommodates a heterogeneous fleet and considers both fixed and operational costs \cite{Hoff2010IndustrialRouting}.

Fleet composition optimization problems are typically posed as a capacitated VRPTW where the fleet size can be varied \cite{Desaulniers2014TheWindows}.
Exact algorithms that guarantee optimality for this combinatorial optimization problem, structure the problem as a tree exploration problem and are solved using Branch \& Bound (B\&B) methods \cite{Desaulniers2014TheWindows}.
However, due to the $\mathcal{NP}$-hard nature of the problem, the application of exact algorithms is restricted to small problem instances \cite{Elshaer2020AVariants}.
Real-life VRPTW applications are considerably larger in scale \cite{Elshaer2020AVariants} and
finding the optimal solution to such a problem is computationally expensive.
Therefore, most VRPTWs are solved using metaheuristic methods due to their ability to find near-optimal solution in a limited time\cite{Elshaer2020AVariants,Desaulniers2014TheWindows}.
However, such approximate methods do not provide guarantees on the optimality of the solution \cite{Desaulniers2014TheWindows}. 

Hybrid optimization methods can improve the performance and efficiency of the optimizer by combining the strengths of metaheuristics and exact algorithms.
Successful metaheuristics provide a balance between exploration and exploitation of the search space \cite{Boussaid2013AMetaheuristics}.
As such, Monte-Carlo Tree Search (MCTS) is a reinforcement learning algorithm that balances this exploration and exploitation
and it is well suited to large-scale combinatorial optimization problems
\cite{Desaulniers2014TheWindows,Silver2016MasteringSearch,Swiechowski2022MonteApplications}. In fact, MCTS has already been used in literature as a metaheuristic that guides a CPLEX solver toward the optimal solution \cite{Sabharwal2012UCT}. Moreover, it is frequently hybridized with other optimization algorithms \cite{Swiechowski2022MonteApplications}.
MCTS has been found to obtain state-of-the-art results in resource allocation problems (RAP) \cite{Kartal2016MonteAllocation} and in single vehicle instances of the VRPTW, called Travelling Salesperson Problems with Time Windows (TSPTW) \cite{Edelkamp2015Monte-CarloLogistics}.
It has also been used to solve VRP problems with variable fleet sizes\cite{Kartal2016MonteAllocation, Edelkamp2015Monte-CarloLogistics, Barletta2022HybridSearch}.
However, MCTS has not yet been used to solve FSMVRPTWs that permit different types of vehicles.


The first contribution of this paper is the development of an exact incremental B\&B algorithm for the FSMVRPTW. 
This algorithm employs a divide and conquer approach where the VRPTW is partitioned into an (RAP) that first assigns tasks to each robot using a parallel B\&B algorithm, and then finds the optimal sequence in which the assigned tasks are completed by solving a nested TSPTW, using another B\&B algorithm. 
The second contribution is a hybrid MCTS-based metaheuristic (UCT-MH), that uses the Upper Confidence bounds applied to Trees algorithm \cite{Kocsis2006BanditPlanning} in the fleet composition levels to guide its search and solves the nested TSPTW using a B\&B algorithm.
The third novelty presented in this paper is the hybrid optimization framework where the UCT-MH guides the incremental B\&B to find the optimal solution to the FSMVRPTW.
When possible, this B\&B is initialized with a fleet composition identified by the rapid search space exploration enabled by the UCT-MH.
Additionally, the best solutions found by the UCT-MH update the upper bound used by the incremental B\&B, which allows sub-optimal solutions to be pruned earlier.
The performance of the proposed method is verified on various real-life case studies. Results show a significant reduction in computation time when the incremental B\&B algorithm is guided by the proposed UCT-MH, especially for large problem sizes.

\section{Problem Formulation \& Methodology}
\label{Methodology}
Consider a manufacturing plant with a known layout that comprises various spatial constraints, and a set of material handling tasks $\mathcal{T}$.
Each task involves picking-up certain cargo items at inventory locations and dropping them off at their designated drop-off locations within defined time windows. 
The objective of the optimization is to find the optimal fleet of multi-load capacitated AMRs that completes all the defined tasks $\mathcal{T}$ while minimizing fixed and operational costs.

Let the set $\mathcal{H} := \{1,2,...,h\}$ identify $h$ different AMR types available, each with specific traveling speeds, energy efficiency, cargo capacity, driving range, charge-time etc. 
Let $k_i\leq k_i^{max}: i \in \mathcal{H}$ denote the number of each type of AMR that forms a fleet so that any fleet composition can be fully defined by a vector $\mathbf{k} \in \mathbb{N}_0^h$.
This fleet is associated with a fixed cost $J^f(\mathbf{k})$ composed of purchase costs, depreciation, etc., that can be captured by $J^f(\mathbf{k}) = \mathbf{c}^\top \mathbf{k}$ for some $\mathbf{c}\in \mathbb{R}^h$.
For completing all the tasks in $\mathcal{T}$, the operational cost $J^o(\mathbf{k})$ can be any combination of relevant metrics to be minimized such as energy, slack time, number of turns, asset depreciation, etc. 
\cite{lu2020time,kucukoglu2021electric,goutham2022}.
The total cost to be minimized is:
\begin{equation}
    \begin{aligned}
        \min_{\mathbf{k}} J = \mathbf{c}^\top \mathbf{k} + J^o(\mathbf{k})\\
    \end{aligned}
\label{eq:Obj1}
\end{equation}

The fleet operational cost $J^o(\mathbf{k})$ is posed as an RAP that finds the optimal partition of tasks to be assigned to AMRs that minimizes total operational cost.
If the total number of robots in the heterogeneous fleet $\mathbf{k}$ is given by $m = \sum_{i=1}^{h} k_i$, every robot in this fleet can be identified by $r\in \mathcal{R}_k:=\{1,2,...,m\}$.
Let the set $\mathcal{T}_r \subseteq \mathcal{T}$ denote the tasks assigned to robot $r$ by the partitioning of $\mathcal{T}$, denoted by $\mathfrak{T}:= \{ \mathcal{T}_r:r \in \mathcal{R}_k \}$, meaning  $\bigcup_{r\in\mathcal{R}_k} \mathcal{T}_r = \mathcal{T}$ and $\forall r,s \in \mathcal{R}_k: r\neq s, \mathcal{T}_r \bigcap \mathcal{T}_s = \varnothing$.
The optimal partition of task set $\mathcal{T}$ minimizes $J^o(\mathbf{k})$ in Eq (\ref{eq:TaskAssignment}).
\begin{equation}
    \begin{aligned}
        J^o(\mathbf{k}) = \min_{\mathfrak{T}} \sum_{r\in\mathcal{R}_k}J^r(r,\mathcal{T}_r)\\
    \end{aligned}
\label{eq:TaskAssignment}
\end{equation}

The minimum operational cost $J^r(r,\mathcal{T}_r)$ for each robot in fleet $\mathbf{k}$ is dependent on the AMR type, and is also affected by the sequence with which task locations are visited, as it is possible for the robot to pickup multiple items before dropping them off so long as each pickup is visited before the corresponding drop-off.
The objective function and constraints that yield $J^r(r,\mathcal{T}_r)$ are defined in Eq. (\ref{eq:obj}-\ref{eq:cargo limit}).


Let robot $r$ of the fleet be assigned $n_r = |\mathcal{T}_r|$ tasks.
The set of pickup and drop-off locations are defined as $\mathcal{V_P}:=\{1,2,...,n_r\}$ and $\mathcal{V_D}:=\{n_r+1,n_r+2,...,2n_r\}$ respectively, so that an item picked up at location $i$ must be dropped off at location $n_r+i$.
The origin and final destination locations of the robot are identified by $\{0,2n_r+1\}$.
Let $\mathcal{V}:=\{\mathcal{V_P} \cup \mathcal{V_D} \cup \{0,2n_r+1\}\}$ be the set of all locations in a graph representation $\mathcal{G}:=(\mathcal{V},\mathcal{A})$ where  $\mathcal{A}:=\{(i,j)\in \mathcal{V}\times\mathcal{V}\}$ is the arc set.
Between every pair of nodes $(i,j)\in \mathcal{A}$, the operational costs $D_{ij}\in\mathbb{R}^+$, energy consumed $\delta e_{ij}\in\mathbb{R}^+$ and travel time $\delta t_{ij}\in \mathbb{R}^+$ are pre-computed before initializing the optimization by solving a path planning problem between every two locations $i,j\in \mathcal{V}:(i,j)\in \mathcal{A}$.
\begingroup
\allowdisplaybreaks
 \begin{align}      
    \label{eq:obj}      &J^r(r,\mathcal{T}_r)  = \min_{ \substack{x_{ij}\\ \forall ij \in \mathcal{A} }}  ~\sum_{i=0}^{2n_r} ~ \sum_{j=1}^{2n_r+1} D_{ij}x_{ij}\\
    \label{eq:binary}   \textrm{s.t.} \quad &x_{ij}\in \{0,1\}                                          ~\forall (i,j)\in \mathcal{A} \\
    \label{eq:depot+}    &\sum_{j = 1}^{n_r} x_{0j}  = 1               \\
    \label{eq:through}  \sum_{i = 0}^{2n_r}& x_{il} = \sum_{j = 1}^{2n_r+1} x_{lj} = 1,      \quad \forall l \in \{\mathcal{V} \setminus \{0,2n_r+1\} \}\\
    \label{eq:depot-}   &\sum_{i = n_r+1}^{2n_r} x_{i,2n_r+1}  = 1                        \\
\label{eq:socEvolution}
 z_j   =& 
  \begin{cases}
   z_i-\delta e_{ij}     \\ ~\; \text{if } x_{ij} =1 \wedge z_i-\delta e_{ij}>0 ~~\forall (i,j)\in \mathcal{A}\\
   1-\delta e_{0j}        \\ ~\; \text{if } x_{ij} =1 \wedge z_i-\delta e_{ij}\leq 0 ~~\forall (i,j)\in \mathcal{A}\\
  \end{cases}
  \\
\label{eq:battery}       \quad z_0 =& 1;~0 \leq z_i \leq 1                           ~\forall i\in \mathcal{V}\\
\label{eq:timeEvolution}
 T_{ij}   =& 
  \begin{cases}
   \delta t_{ij}       \\ \quad  ~  \text{if } x_{ij} =1 \wedge z_i-\delta e_{ij}>0  ~~\forall (i,j)\in \mathcal{A}\\
   \delta t_{0i}+(1-z_i-\delta e_{i0})p^{-1}+\delta t_{0j}      \\ \quad ~  \text{if } x_{ij} =1 \wedge z_i-\delta e_{ij}\leq 0 ~~\forall (i,j)\in \mathcal{A}
  \end{cases}\\
    \label{eq:travel time}      &x_{ij}=1 \rightarrow t_i+s_i+T_{ij}\leq t_{j}      ~\forall (i,j)\in \mathcal{A}\\
    \label{eq:cut}              & \quad t_i +s_i + T_{i,n+i} \leq t_{n+i}                   ~\forall i\in \mathcal{V}\\
    \label{eq:time window}      & \quad  e_i \leq t_i \leq l_i                   ~\forall i\in \mathcal{V}\\
    \label{eq:cargo update}     &x_{ij} =1 \rightarrow y_j = y_i + q_j           ~\forall (i,j)\in \mathcal{A}\\
    \label{eq:cargo limit}      & \quad y_0 = 0;~0 \leq y_i \leq Q_r                           ~\forall i\in \mathcal{V}
\end{align}
The binary flow variable $x_{ij} = 1$ signifies that the robot uses directed arc $(i,j)\in \mathcal{A}$.
Constraints related to the robot starting from the depot $0$, visiting every location once and terminating the sequence at $2n_r+1$ are enforced by Eq. (\ref{eq:binary}-\ref{eq:depot-}).

The battery states of charge $z_j$ in Eq. (\ref{eq:socEvolution}-\ref{eq:timeEvolution}) are updated as the robot goes about its mission.
Whenever the battery is depleted, the robot heads to the depot where it is fully charged up with a constant recharging rate $p$.
The variable $T_{ij}$ in Eq. (\ref{eq:timeEvolution}-\ref{eq:cut}) updates the travel time between locations $i$ and $j$ based on whether a recharge is required between the two locations.
Time variables $t_i$ in Eq. (\ref{eq:travel time}-\ref{eq:time window}) denote the arrival time of the robot at location $i \in \mathcal{V}$.
Each location  is associated with a time $s_i$ for material handling and a time window $[e_i,l_i]$ which represents the earliest and latest time at which material handling can start.
Cargo constraints are captured in Eq. (\ref{eq:cargo update}, \ref{eq:cargo limit}) where payload variables $y_i$ capture the cargo mass being carried by the robot as it leaves location $i\in \mathcal{V}$.
Each robot $r$ has a cargo capacity limitation of $Q_r$ and each location $i\in\mathcal{V}$ is associated with a cargo load $q_i\in\mathbb{R}$ such that $q_i+q_{n+i}=0$.
 Volumetric constraints are modeled similarly. 

\subsection{Exact Algorithm: Incremental Branch \& Bound}\label{exactAlgorithm}
The incremental B\&B systematically partitions the search space into subsets that are arranged in a tree structure.
The root of the tree is the original problem and the leaves of the tree are its individual candidate solutions.
Between the root and the leaves are intermediate nodes that represent subproblems obtained by 
recursively partitioning the original problem by a process called branching. B\&B algorithms are used to solve these sub-problems.
The order according to which these subproblems are examined is determined by a best-first selection criteria, i.e. exploitation, that first explores the problem with the cheapest cost.

For minimization problems, the upper bound is the incumbent solution which is the cheapest candidate solution to the original problem found at the leaf node.
The upper bound is continuously updated as the tree is explored, and is used to prune sub-optimal branches without recursively evaluating their solutions up to the leaf node.
Thus, as the algorithm searches from the root to the leaves, branching is conducted only if the cost at the node is lower than the incumbent solution, and branching can potentially find a better solution than the incumbent solution.
Following this process, the B\&B algorithm recursively decomposes the original problem until further branching is futile when the solution cannot be improved, or until the original problem has been solved when every feasible branch has been evaluated.


\begin{figure}[b]
    \centering
        \includegraphics[trim =8mm 35mm 30mm 3mm, clip, width=0.99\linewidth]{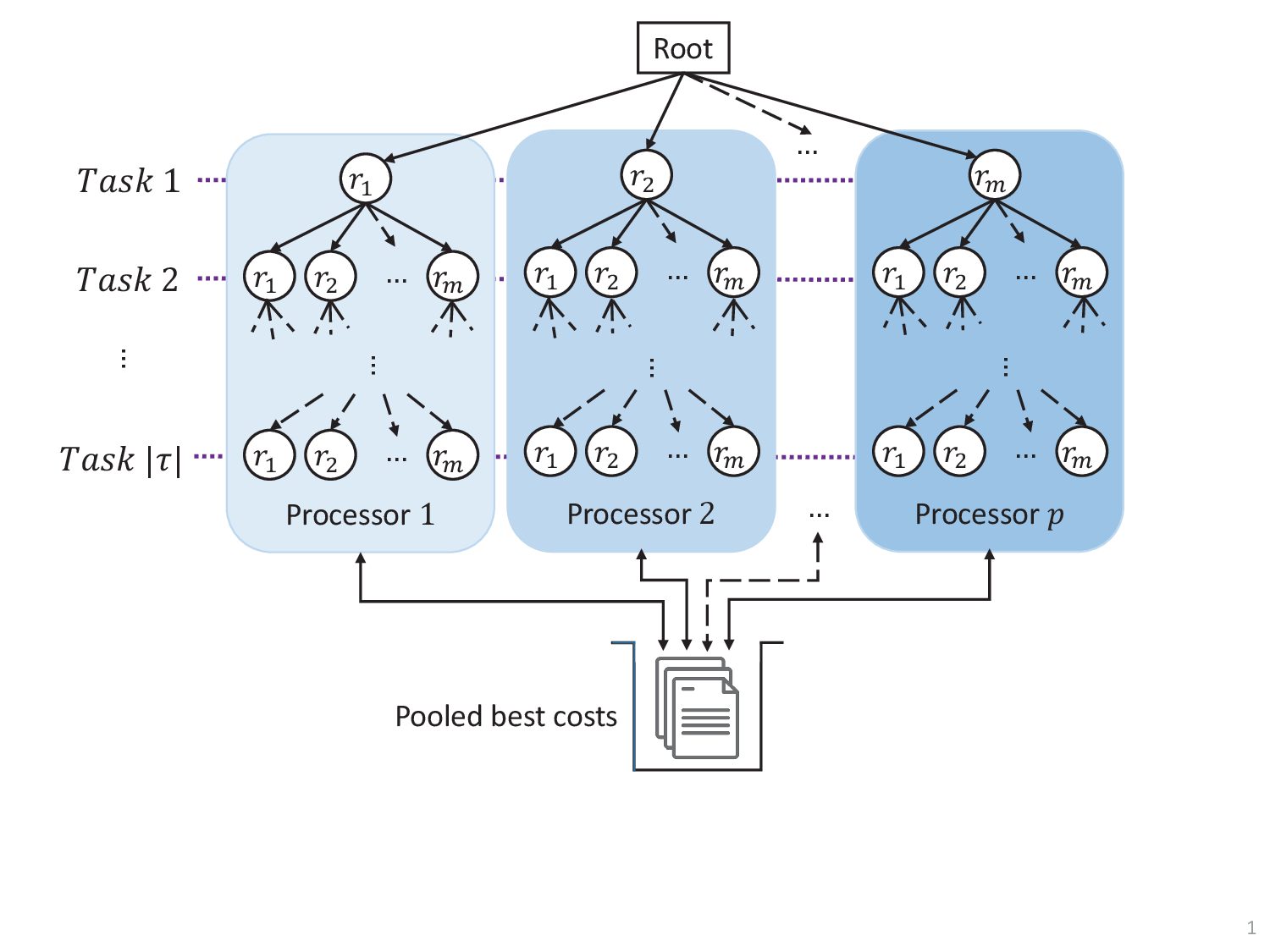}
  \caption{Parallel implementation of the RAP B\&B algorithm for the Resource Allocation Problem formulation.}
  \label{fig:ParallelTA} 
\end{figure}

The $\mathcal{NP}$-hard RAP problem described by Eq. \ref{eq:TaskAssignment} is solved by the B\&B algorithm implemented in a parallel framework that uses $p$ processing cores, as shown in Fig. \ref{fig:ParallelTA}, where robots in the fleet are identified by subscript $r\in \mathcal{R}:=\{1,2,...,m\}$.
Thus, by splitting the arborescence at some task assignment level and assigning the emanating sub-trees to the available processors, several subproblems are explored simultaneously.
During each processor's exploration, updated incumbent solutions are instantaneously made available to every processor in an asynchronous information sharing method using a shared work pool.
For each processor, this RAP B\&B algorithm is implemented by a recursive function to minimize memory and computational requirements as the tree is explored.
Further, since the computation time of B\&B algorithms increases with the number of feasible branches at each node, the fleet is initiated with a smaller candidate fleet $\mathbf{f}^1\in \mathbb{N}_0^h$ than the maximal fleet $\mathbf{k}^{max} \in \mathbb{N}_0^h$.
After evaluating the total cost of this candidate fleet, the number of robots is incrementally raised until further increments do not reduce the total cost or additional robots remain idle.
For each fleet increment, only RAP subproblems that include at least one of the newly added robots are evaluated since other solutions are guaranteed to have been evaluated already.

For $h$ different AMR types available, the fleet is initiated with a candidate fleet $\mathbf{f}^1\leq\mathbf{k}^{max}$, which is chosen based on problem parameters and prior experience so that feasible solutions exist.
The RAP of fleet $\mathbf{f}^1$ is then solved using the described parallel B\&B algorithm, and its minimum total cost $J^1$ is found, which utilizes robots $\mathbf{k}^1\in \mathbb{N}_0^h :\mathbf{k}^1 \leq \mathbf{f}^1$.
In the increment step, only robot types $i$ that satisfy $\mathbf{k}^1_i = \mathbf{f}^1_i$ are incremented by 1 for the next candidate fleet $\mathbf{f}^2$.
These increments are conducted so long as both $J^{i+1}\leq J^i$ and $\exists ~ i: \mathbf{k}^1_i = \mathbf{f}^1_i$.
The optimal fleet that completes all tasks while minimizing total cost is then $k^{i}$ when $J^{i+1}>J^i$ or when $\nexists ~ i: \mathbf{k}^{i+1}_i = \mathbf{f}^{i+1}_i$, i.e, the additional robots were idle.

At each instance that the RAP subproblem is solved at a node in the arborescence shown in Fig. \ref{fig:ParallelTA}, the TSPTW problem defined by Eq. (\ref{eq:obj}-\ref{eq:cargo limit}) is solved to find the cost at that node.
This TSPTW problem is solved using recursive Algorithm \ref{algorithm1} that employs another B\&B to find the optimal sequence of task completion for each robot. 
In summary, the B\&B incrementally increases the fleet size while minimizing total cost $J$ of Eq. \ref{eq:Obj1},
which includes operational cost $J^o(\mathbf{k})$  of Eq. \ref{eq:TaskAssignment} found using the RAP B\&B algorithm
and the cost $J^r(r,\mathcal{T}_r)$ of Eq. (\ref{eq:obj}-\ref{eq:cargo limit}) found using the TSPTW B\&B algorithm.

\begin{algorithm}
\caption{Recursive TSPTW B\&B\\
Cost  = \emph{B\&B}(State, taskList, current Location)}\label{algorithm1}
\begin{algorithmic}[1]
\STATE Find feasible next locations based on completed pickups, time, cargo, battery constraints
\STATE Sort feasible next locations by operational cost of branching to that location (Best First Search)
\FOR{\emph{i} in feasibleLocations}
    \STATE \emph{branchCost} = tourCost + operationalCost(i)
    \IF{\emph{branchCost} $\geq$ State.bestCost}
        \STATE continue \{ skip to next \emph{location i+}\}
    \ELSIF{\emph{branchCost}$<$ State.bestCost}
        \STATE \emph{State+} = Update stateOfTime, stateOfCharge, finalPosition, remainingLocations
            \IF{remainingLocations $> 0$}
                \STATE Cost = \emph{B\&B}(State+, taskList, \emph{location}(i))
            \ELSE
                \STATE State.bestCost = Cost
            \ENDIF
    \ENDIF
\ENDFOR
\end{algorithmic}
\end{algorithm}

\subsection{Metaheuristic: Monte-Carlo Tree Search}
Each iteration of MCTS involves four steps \cite{Browne2012AMethods}:
\textit{Selection:} at every node $v$ in the arborescence, the tree policy selects the next node $v'$. This node selection is initiated at the root node $v_0$ and is used for navigation until the leaf node $v_l$ is reached. 
\textit{Expansion:} at the leaf node $v_l$, a random action is taken to expand the tree.
\textit{Simulation:} a Monte-Carlo simulation is performed starting from the expansion node to complete the solution. 
\textit{Backpropagation:} the cost/reward of the expansion and simulation is propagated back to the root node $v_0$.

The Upper Confidence bounds applied to Trees algorithm \cite{Kocsis2006BanditPlanning} was the first variant and formal introduction of MCTS.
The proposed metaheuristic (UCT-MH) uses this algorithm to guide the exact incremental B\&B algorithm to the optimal solution.
While in typical VRPTWs the fleet size is a free variable \cite{Desaulniers2014TheWindows}, the proposed metaheuristic selects a fleet size $m$ and a composition $\mathbf{k}$ in the fleet sizing and composition stages and tries to solve the VRPTW optimally, fully utilizing that composition. 
By doing so, the algorithm finds an estimate of the expected total cost associated with a particular fleet size and composition. 
This estimate serves as a measure for the quality of that branch and can be used by the MCTS to navigate the search. 
MCTS is most effective as a heuristic at the early stages of the decision problem \cite{Sabharwal2012UCT}.
Moreover, for smaller problem instances, B\&B algorithms are often more suitable than MCTS \cite{Edelkamp2015Monte-CarloLogistics}.
As such, the proposed hybrid MCTS algorithm is aimed to utilize the strengths of the different algorithms and combine them into an effective hybrid MCTS-based metaheuristic.

\begin{figure}[b]
    \centering
\includegraphics[width=0.85\columnwidth]{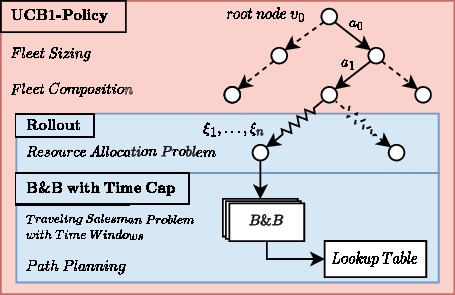}
    \caption{Overview of the multi-stage design problem, with the FSMVRPTW (red) and the nested VRPTW (blue), and the proposed UCT-MH Algorithm.}
    \label{fig:UCT-MH-Schematic}
\end{figure}

Although MCTS was originally designed to solve Markov Decision Processes, without loss of generality, MCTS can be used to solve a design problem by formulating it as a deterministic Markov Decision Process \cite{Swiechowski2022MonteApplications}. 
The optimization problem is modeled as a 3-tuple $\langle S,A,g \rangle$, where $S$ is a set of states, $A$ is a set of actions and $g(s,a): S \times A \rightarrow [0, g_{max}]$ is a scalar cost function for taking action $a$ at state $s$. The state $s(v)$ contains the parameters that follow from the decisions up to node $v$.
At the root node $v_0$, the fleet size $m$ is determined by action $a_0$, where $g_1(s_0(v_0), a_0) := 0$, for the fleet cost is determined by its composition. Subsequently, the fleet composition $\mathbf{k}$ is determined by $a_1 \in A_1(m)$, with fixed cost $g_2(s_1(m),a_1) = J^f(\mathbf{k})$.
Fig. \ref{fig:UCT-MH-Schematic} provides a schematic overview of the problem and the proposed metaheuristic. 

At the fleet sizing and composition stages, the UCT-MH utilizes the UCB1 tree policy \cite{Kocsis2006BanditPlanning} for the selection step at node $v$ of the search tree:
\begin{equation}
    \text{UCB1}(v) = \operatorname*{arg\,max}_{v' \in \text{children of } v} \frac{Q(v')}{N(v')} + \sqrt{\frac{2 \ln N(v)}{N(v')}}
    \label{eq:UCB}
\end{equation}
Here, $Q(v')$ is the total reward of all plays through child node $v'$, $N(v')$ denotes the number of visits of child node $v'$, and $N(v)$ is the number of visits of the parent node $v$. The policy function is dependent on the quality of the node being considered as well as the number of evaluations of that node, balancing the exploration and exploitation of the search space\cite{Browne2012AMethods}.
In order to apply the UCB1 policy and have a proper balance between exploration and exploitation, the problem is transformed such that the stage reward $R_i(v) \in [0, 1]$ \cite{Kocsis2006BanditPlanning}:
\begin{equation}
R_i(v') = 1 - \frac{g_i(v')}{g_{max}}
\end{equation}
where $R_i(v')$ is the reward of the transition from state $s_{i-1}(v)$ to state $s_{i}(v')$ and $v'\in \text{children of } v$.
It follows that $Q(v')$ is the sum of all rewards of all $N(v')$ plays through node $v'$ back to the root node $v_0$:
\begin{equation}
    Q(v') = \sum_{i=1}^{N(v')} R_i(v') + R_i(v) + ... + R_i(v_0)
\end{equation}

Considering that the number of permutations of the RAP is exponential with the number of tasks, it is deemed sufficient to determine the task assignment by a random rollout $\left( \xi_1, ..., \xi_n \right)$. 
In order to prevent any bias toward another fleet size, it is ensured that the full fleet size is utilized, i.e. each AMR in the fleet will have at least one assignment.
The assigned tasks do not have any associated costs/rewards.

Since many of the TSPTW instances encountered are small problem instances, it is advantageous to use the same recursive B\&B algorithm for TSPTW as described in Section \ref{exactAlgorithm} to find the optimal sequence in which the assigned tasks are completed by each robot.
Each TSPTW B\&B is terminated after a one second time cap since the metaheuristic is not aimed at local convergence.
Considering the best first order of exploration, this still finds reasonably good estimates for the operational cost $\Tilde{J}^o(k)$.
The cost that is obtained through the rollout of the RAP and the TSPTW, is backpropagated through the tree and are assigned to $Q(v)$ at node $v$ that is associated with a particular fleet size or composition. This is in turn used by the UCB1 policy function to determine the decisions in the next iteration.
As a result, at the root node, the term $\frac{Q(v')}{N(v')}$ in (\ref{eq:UCB}) is proportional to the total mean cost-to-go for a given fleet size or composition at node $v'$. As the total number of plays at the root node $N(v_0)$ grows to infinity, the UCB1 function converges to the expected value of the total cost for a given fleet size.

\subsection{Hybrid Optimization: Guiding B\&B with the UCT-MH} \label{Parallel MCTS}
The hybrid optimization framework utilizes the search results of the UCT-MH to guide the exact incremental B\&B. Multiple processors are allocated to the B\&B algorithm that systematically navigates the tree to solve the problem exactly. 
Meanwhile, one processor is dedicated to running the UCT-MH which efficiently samples the entire design space to get an estimate of the associated costs. 
Considering the parallelization overhead of the paralleled B\&B algorithm, it can be expected that the UCT-MH already finds a fleet composition candidate $\mathbf{\hat{k}}$ by the time the B\&B is initiated.
If such a composition is available, then it is used as the candidate fleet $\mathbf{f}^1=\mathbf{\hat{k}}$ that initializes the B\&B algorithm.
Moreover, whenever the guiding UCT-MH finds a new best solution, it provides this solution with its associated cost to the guided B\&B by adding it to the pooled best cost shown in Fig. \ref{fig:ParallelTA}. 
This information is used to preemptively prune sub-optimal branches and guide the B\&B toward the optimal fleet size and composition, thereby reducing the search space and computation time.

\section{Results}
\label{Results}
\begin{figure}[t!]
    \centering
    \vspace{-1mm}
   \subfloat[The quality of candidate fleet sizes as determined by the UCT-MH \label{a:10-6}]{%
        \hspace{2.38mm}\includegraphics[trim =0mm 0mm 0mm 0mm, clip, width=0.975\linewidth]{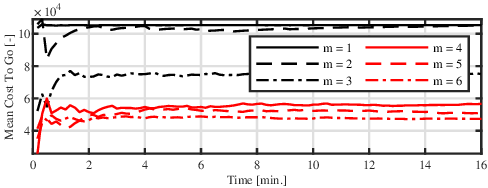}}\\
        \vspace{-3mm}
    \subfloat[The UCT-MH balances exploration and exploitation of candidate solutions. \label{b:10-6}]{%
        \includegraphics[trim =0mm 0mm 0mm 0mm, clip, width=1.0\linewidth]{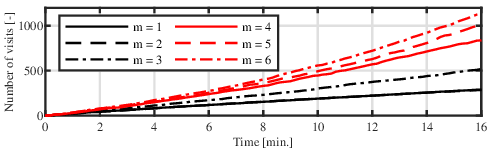}}\\
        \vspace{-6mm}
   \subfloat[Performance of UCT-MH and B\&B algorithm and their parallelization. \label{c:10-6}]{%
        \hspace{1.92mm}
        \includegraphics[trim =0mm 0mm 0mm 0mm, clip, width=0.962\linewidth]{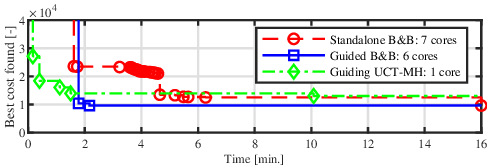}}
  \caption{Case Study 1: The UCT-MH and B\&B algorithm, number of tasks $n=10$, maximum number of AMRs: $m_{max} = 6$, $\mathbf{k}_{max}^\top = [2,2,2]$.}
  \label{fig:10-6} 
     \vspace{3mm}
     \subfloat[The quality of candidate fleet sizes as determined by the UCT-MH, first 60 minutes of simulation.\label{a:20-21}]{%
        \includegraphics[trim =0mm 0mm 0mm 0mm, clip, width=1\linewidth]{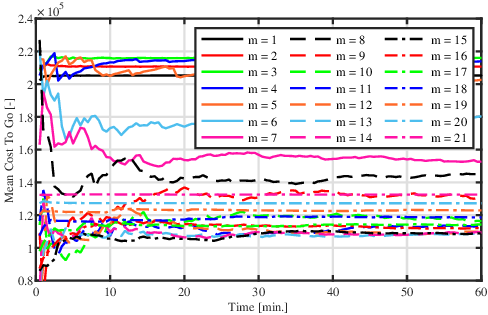}}\\
        \vspace{-3mm}
   \subfloat[Performance of UCT-MH and B\&B algorithm and their parallelization.
   \label{b:20-21}]{
        \includegraphics[trim =0mm 0mm 0mm 0mm, clip, width=1\linewidth]{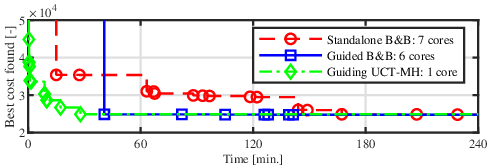}}
  \caption{Case Study 2: The UCT-MH and B\&B algorithm, number of tasks $n=20$, maximum number of AMRs: $m_{max} = 21$, $\mathbf{k}_{max}^\top = [7,7,7]$.}
  \label{fig:20-21}
\end{figure}

\subsection{Computational Experiments}
To study the performance of the proposed hybrid algorithm, the guiding UCT-MH and the guided B\&B are compared against the standalone incremental B\&B.
Four real-life case studies are conducted in MATLAB 2022a at the Ohio Supercomputer Center \cite{OhioSupercomputerCenter1987}. 
For each experiment, a set of $n$ tasks is defined, each consisting of items of known mass, volume, pick-up and drop-off locations and respective time windows. 
The fleet size is limited to $m_{max}$, equally distributed over $h=3$ different AMR types. Each algorithm is run for a limited time $t_{max}$ after which the incumbent solutions are compared. Two smaller problems are studied in detail to illustrate the behavior of the UCT-MH in Fig. \ref{fig:10-6}-\ref{fig:20-21}. 
The best-found cost by each algorithm is summarized for all case studies in Table \ref{tab1}.

\subsection{Case Studies}
\subsubsection{$n=10$ and $m_{max} = 6$}
Figure \ref{a:10-6} shows the UCT-MH exploration of the various fleet sizes, where the mean of the cost-to-go starts to converge and the algorithm gains more confidence in particular solutions as the number of evaluations increases.
The guiding UCT-MH finds that $m=6$ is the best candidate and dedicates more visits to these branches as shown in Fig. \ref{b:10-6}.
As a result, the guided B\&B quickly focuses on local convergence (Fig. \ref{c:10-6}). As the entire search space is explored, this solution is the guaranteed global optimum.

\subsubsection{$n=20$ and $m_{max} = 21$}
In Fig. \ref{a:20-21} several patterns are observed. While small fleet sizes yield infeasible solutions, larger fleet sizes initially show a transient behavior due to the stochastic exploration. 
The largest fleet sizes always yield feasible solutions, irrespective of the lower-level decisions.
Here, an increase in fleet size results in an incremental increase of the mean cost-to-go which is associated with the fleet cost.
Remarkably, Fig. \ref{b:20-21} shows that the standalone B\&B is initially faster, however, as the guided B\&B already starts from a good candidate branch, the underlying TSPTW is expected to be more difficult to solve. 
Consequently, the guided B\&B discards suboptimal fleets and focuses on local convergence thereby reducing the overall computation time of the guided B\&B.

\subsection{Discussion}
The time taken to initialize the parallel B\&B algorithm is sufficient for the guiding UCT-MH to find a strong candidate fleet that warm starts the guided B\&B.
The UCT-MH provides a reduction of computation time ranging from $38.3\%$ up to $86.5\%$.
The local convergence of the UCT-MH is dependent on the problem size due to the time cap imposed at the TSPTW level. 
As seen in Table \ref{tab1}, for a higher number of tasks where the TSPTW is larger, the gap with the best-known solution is greater ($\sim \hspace{-0.55mm}40\%$).
However, the guided B\&B is able to close this gap since it conducts local searches systematically.
Further, for the case with 100 tasks, the standalone B\&B was unable to find any feasible solution in 24 hours while the UCT-MH provided multiple solutions through its efficient stochastic exploration of the design space.

\begin{table*}[t]
\vspace{-1mm}
\caption{Experimental Results - B\&B and UCT-MH}
\begin{center}
\renewcommand{\arraystretch}{1.1}
\vspace{-2.7mm}
\begin{tabular}{|ccc|rr|rrrr|rrr|ll}
\cline{1-12}
\multicolumn{3}{|c|}{\textbf{Case Study}} &
  \multicolumn{2}{c|}{\textit{\textbf{Standalone B\&B}}} &
  \multicolumn{4}{c|}{\textit{\textbf{Guided B\&B}}} &  \multicolumn{3}{c|}{\textit{\textbf{Guiding UCT-MH}}} &
   &
   \\ \cline{1-12}
\multicolumn{1}{|c|}{$\mathbf{n}$} &
  \multicolumn{1}{|c|}{$\mathbf{m}_{\text{max}}$}&
  \textbf{t$_{\text{max}}$ [h]} &
  \multicolumn{1}{c|}{\textbf{Cost}} &
  \textbf{t$_{\text{found}}$ [min]} &
  \multicolumn{1}{c|}{\textbf{Cost}} &
  \multicolumn{1}{c|}{\textbf{Rel. Gap}} &
  \multicolumn{1}{c|}{\textbf{t$_{\text{found}}$ [min]}} &
  \textbf{Reduction} &
  \multicolumn{1}{c|}{\textbf{Cost}} &
  \multicolumn{1}{c|}{\textbf{Rel. Gap}} &
  \textbf{t$_{\text{found}}$ [min]} &
   &
   \\ \cline{1-12}
\multicolumn{1}{|c|}{10} &
\multicolumn{1}{|c|}{6} &
  2 &
  \multicolumn{1}{r|}{9633.5$^*$} &
  16 &
  \multicolumn{1}{r|}{9633.5$^*$} &
  \multicolumn{1}{r|}{0.00\%} &
  \multicolumn{1}{r|}{2.17} &
  86.5\% &
  \multicolumn{1}{r|}{10190.8} &
  \multicolumn{1}{r|}{5.79\%} &
  78.52 &
   &
   \\ \cline{1-12}
\multicolumn{1}{|c|}{20} &
  \multicolumn{1}{|c|}{21} &
  4 &
  \multicolumn{1}{r|}{24,797.0} &
  229.02 &
  \multicolumn{1}{r|}{24,761} &
  \multicolumn{1}{r|}{- 0.15\%} &
  \multicolumn{1}{r|}{141.35} &
   38.3\% &
  \multicolumn{1}{r|}{24,891.5} &
  \multicolumn{1}{r|}{0.53\% } &
  27.88 &
   &
   \\ \cline{1-12}
\multicolumn{1}{|c|}{50} &
\multicolumn{1}{|c|}{30} &
  12 &
  \multicolumn{1}{r|}{38,047.3} &
  420.43 &
  \multicolumn{1}{r|}{40,638.3} &
  \multicolumn{1}{r|}{+ 6.81\%} &
  \multicolumn{1}{r|}{212.30} &
   49.5\% &
  \multicolumn{1}{r|}{53,161.3} &
  \multicolumn{1}{r|}{39.72\%} &
  \multicolumn{1}{r|}{2.883} &
   &
   \\ \cline{1-12}
\multicolumn{1}{|c|}{100} &
\multicolumn{1}{|c|}{60} &
  24 &
  \multicolumn{1}{|r|}{N/A$^\dag$} &
 $-$ &
  \multicolumn{1}{r|}{74,250.5} &
  \multicolumn{1}{r|}{$-$} &
  \multicolumn{1}{r|}{358.33} &
  $-$ &
  \multicolumn{1}{r|}{103,193.0} &
  \multicolumn{1}{r|}{38.98\%} &
  30.05 &
   &
   \\ \cline{1-12}
\end{tabular}

\label{tab1}
\end{center}
\text{$^*$Globally optimal solution.}\\
\text{$^\dag$No solution was found after 24 hours.}
\end{table*}



\section{Conclusions}
\label{Conclusion}
In this paper, a hybrid optimization algorithm was developed that uses a Monte-Carlo Tree Search-based metaheuristic (UCT-MH) to guide an exact incremental Branch \& Bound algorithm, which solves a real-life Fleet Size and Mix Vehicle Routing Problem with Time Windows. 
The UCT-MH yields a significant improvement in the computation time and convergence of the B\&B by constantly sharing the expected optimal fleet composition as well as the upper bound on the cost. 
Although in this study MCTS was only employed at the fleet sizing and composition level, future research needs to determine to what depth MCTS can be effective. Moreover, modifications to the selection policy as well as bi-directional communication between the UCT-MH and the B\&B algorithm could further improve computation times.

\section*{Acknowledgments}
This research was supported by the Ford Motor Company as part of the Ford-OSU Alliance Program.

\medskip
\bibliographystyle{IEEEtran}
\bibliography{manualRef}

\end{document}